\documentclass[prd,onecolumn,notitlepage,showpacs,preprintnumbers,amsmath,amssymb,nofootinbib,APS,10pt,superscriptaddress]{revtex4-1}

\usepackage{dcolumn}
\usepackage{bm}
\usepackage{xcolor}
\usepackage[applemac]{inputenc}
\usepackage[spanish,english]{babel}
\usepackage{amsmath,amssymb,amsfonts,latexsym,cancel}
\usepackage{graphicx}
\usepackage{color}
\usepackage{soul}
\usepackage{ulem}
\usepackage{hyperref}
\usepackage{amsmath}
\usepackage{slashed}
\usepackage{float}

\allowdisplaybreaks

\newcommand{\be}{\begin{equation}}
\newcommand{\ee}{\end{equation}}
\newcommand{\bea}{\begin{eqnarray}}
\newcommand{\eea}{\end{eqnarray}}

\begin{document}

\title{{\bf Role of gravity in the pair creation induced by electric fields} }

\author{Antonio Ferreiro}\email{antonio.ferreiro@ific.uv.es}
\author{Jose Navarro-Salas}\email{jnavarro@ific.uv.es}
\author{Silvia Pla}\email{silvia.pla@uv.es}

\affiliation{Departamento de Fisica Teorica and IFIC, Centro Mixto Universidad de Valencia-CSIC. Facultad de Fisica, Universidad de Valencia, Burjassot-46100, Valencia, Spain.}

\begin{abstract}

We analyze the pair production induced by homogenous, time-dependent electric fields in an expanding space-time background.
We point out that, in obtaining the semiclassical Maxwell equations,  two distinct notions of adiabatic  renormalization are possible. In Minkowski space the two recipes turn out to be  equivalent. However, in the presence of gravity only the recipe requiring an adiabatic   hierarchy between the gravitational and the gauge field is consistent with the conservation of the energy-momentum tensor. \\



\end{abstract}


\date{\today}
\maketitle

\section{Introduction}

The relativistic Dirac theory predicted the existence of antiparticles and the possibility of particle creation and annihilation if the particles  are described by  quantum field theory. 
The presence of a time-dependent gravitation field, as that describing the expanding Universe,  permits the spontaneous creation of particles out of the vacuum \cite{parker66, parker68}. In the latter case, the underlying mechanism is simple:  creation operators of quantized fields living in an expanding universe evolve into a superposition of creation and annihilation operators. 
This breakthrough had a major impact in the understanding of the quantum behavior of black holes  \cite{hawking74}, the physics of the very early Universe \cite{ford-parker}, and subsequent research in quantum aspects of gravitational physics. Similar behavior in the evolution of the creation operators is obtained when the background field is a time-varying electric field, as pointed out in Ref. \cite{BI}, thus recovering in this way the Schwinger effect in the limit of a constant electric field \cite{Schwinger51}. In this paper, we will mainly focus on the vacuum pair production induced by electric fields. Since this effect could be on the verge of being experimentally verified in laboratories \cite{ELI} it is therefore very natural to scrutinize the theoretical side of the phenomena \cite{Dunne, Pittrich-Gies}. The pair creation in electric fields is also important in astrophysics and cosmology \cite{ruffini}. \\

An important problem is the backreaction of the background field due to the created pairs. One faces here a  basic important problem, both conceptually and technically. The source fields in both semiclassical Maxwell and Einstein equations are vacuum expectation values of composite operators, namely, the electric current $\langle j^\mu \rangle$ and the stress-energy tensor $\langle T^{\mu\nu} \rangle$, respectively. The source operators suffer from the typical ultraviolet divergences of composite operators in quantum field theory, and one needs to find consistent ways to renormalize them. Since the basic origin of particle creation is the existence of time-varying backgrounds it is natural to attack the problem by considering first spatially homogeneous backgrounds. In these scenarios, which are extremely important in cosmology,  we have a very sound and efficient renormalization scheme: the adiabatic regularization method. It was originally introduced to bypass the infinities that arose in the particle number during the expansion of the Universe. It was later refined to construct well-defined stress-tensor operators of scalar fields in expanding universes \cite{parker-fulling}. (For Dirac fields see the recent works \cite{rio1}). In the early 1990s, it was used as a basic tool to study pair production in strong electric fields in Minkowski space \cite{Cooper1,Kluger91,Kluger92,Kluger92bis}. In most of the literature on this latter problem, it is assumed a specific implementation of the adiabatic renormalization program without realizing an underlying inconsistency when gravity is turned on. This fact has been largely overlooked in the literature, and we devote this paper to clarifying this issue. \\

\section{Model and the adiabatic renormalization scheme}

To focus on basic ideas, we will consider a simple prototype model, namely, two-dimensional scalar QED in an expanding metric of the form $ds^2 = dt^2 - a^2(t)dx^2$. The discussion can be extended to higher dimensions and for spinor QED.
The classical action of our model is given  by
\bea
\mathcal{S}=\int d^2x \sqrt{-g}\left(-\frac14 F_{\mu\nu}F^{\mu\nu}+(D_{\mu}\phi)^{\dagger}D^{\mu}\phi-m^2 \phi^{\dagger}\phi \right ).
\eea
The quantized scalar field $\phi$ obeys the Klein-Gordon field equation
\be (D_\mu D^\mu + m^2)\phi =0 \ , \ee
where
\be D_\mu \phi = (\nabla_\mu + iqA_\mu )\phi . \ee
We assume  that the gauge field is spatially homogeneous and only time dependent. 
For our purposes it is very convenient to choose a gauge such that only the $x$ component of the vector potential is nonvanishing: $A_{\mu}= (0, -A(t))$. Therefore, the field strength  is  given by $F_{01}(t)=E(t) a(t)= -\dot A(t)$.
The spatial homogeneity of the background metric and the vector potential imply the following Fourier expansion of the quantized field, 
\be \phi(x)= \frac{1}{\sqrt{2(2 \pi a)}}\int_{-\infty}^{\infty} dk [A_{k}e^{ikx}h_{k}(t)+B_{k}^{\dagger}e^{-ikx}h^*_{- k}(t) ] \ , \label{phisolution2} \ee
where
 $A_{{k}}^{\dagger}, B_{{k}}^{\dagger}$ and $A_{{k}}, B_{{k}}$ are the usual creation and annihilation operators, obeying the usual commutation rules $[A_k,A_{k'}^{\dagger}]=\delta (k-k')$ and $[B_k,B^{\dagger}_{k'}]=\delta(k-k')$. 
The commutation rules demand that the mode functions $h_{k}(t)$ obey the Wronskian  condition
$h_{ k}\dot h_{ k}^* - h_{ k}^*\dot h_{ k} = 2i$.
The equation for the functions $h_k(t)$ can be derived from the original field equations, and it reads as 
\be
\ddot{h}_{k}+\left(m^2+\frac {1}{a^2}\left(k-qA\right)^2+ \frac{\dot a^2}{4a^2} -\frac{\ddot a}{2a}\right)h_{k}=0  \ . \label{equhk2} \ee \\
The classical electric current is given by\\
\bea
j^{\nu}=iq\left[\phi^{\dagger}D^{\nu}\phi-(D^{\nu}\phi)^{\dagger}\phi \right] \nonumber\ ,
\eea
and the formal expression for the vacuum expectation values of $\langle j^x \rangle$ is 
\bea
\langle j^{x}\rangle=q \int_{-\infty}^{\infty}\frac{dk}{2\pi a^3} \left(k-q A\right) |h_k|^2. \\
\eea
The above expression possesses, as expected,  UV divergences. To obtain the finite, physical values, we have to perform appropriate subtractions
\be \langle j^{x}\rangle_{ren}=q \int_{-\infty}^{\infty} \frac{dk}{2 \pi a^3} \left [\left(k-q A\right) |h_k|^2- SUBTRACTIONS \right ] \ . \ee 
The semiclassical Maxwell equations are
\bea
\nabla_{\mu}F^{\mu\nu}=\langle j^{\nu} \rangle_{ren} \label{semimax}  \ . \eea
Equations \eqref{equhk2} and \eqref{semimax} determine completely the continuous interchange of energy between the electric field and matter, via charged pair production and backreaction.\\




The main problem we have to discuss is how to obtain the required subtractions in a physically consistent way. As mentioned above, one can determine  the renormalization subtractions from the adiabatic regularization method. The basic principles can be borrowed from Ref. \cite{parker-toms}.
The main idea in dealing with scalar fields is to consider an adiabatic expansion of the mode function $h_k(t)$ based on the WKB-type ansatz , namely
\be h_{k}(t) =\frac{1}{\sqrt{\Omega_k(t)}}e^{-i \int^{t}\Omega_k(t')dt'} \ , \hspace{0.5cm}  \Omega_k(t) = \omega^{(0)}_k + \omega_k^{(1)} + \omega_k^{(2)} + \cdots \, \ee
where the order of the expansion is based on the number of  derivatives of the background fields. A very crucial point to properly define the adiabatic expansion is to define the leading order, i.e., the  zeroth order adiabatic term. In other words, we have to define $ \omega^{(0)}_k$. \\

\section{Two attempts toward the backreaction equations in Minkowski space}
Let us first assume for simplicity that we are in Minkowski space and $a(t)=1$. Therefore, the gauge field is now the only background field. 

\subsection{First choice for $\omega^{(0)}_k$}

In this situation it is very natural to define 
\be \label{1omega}\omega^{(0)}_k = \sqrt{(k-qA)^2 + m^2} \ , \ee
as first assumed in Refs. \cite{Cooper1,Kluger91,Kluger92,Kluger92bis} and in all subsequent papers on this topic. This means that the adiabatic order assignment for the gauge field $A(t)$ has been implicitly chosen as $0$. Therefore, $\dot A(t)$ should be of order $1$, etc.  This rule is displayed in  Table \ref{table1}.
\begin{table}[h]
    \begin{tabular}{ |l|c| }
    \hline
       Field &  \small {Adiabatic order assignment}   \\ \hline
       $A(t)$ &  0  \\   \hline
       $\dot  A(t)$  &  1  \\ \hline
    $\ddot A(t)$, $\dot A^2(t)$,... &  2  \\ \hline
    $ \dddot A(t) $, $\dot A(t) \ddot A(t)$,... &  3  \\ \hline    \end{tabular}
\caption{\small{We  summarize the adiabatic order assignment for different derivatives of the gauge field according to the assumptions in the previous literature.  }}
\label{table1}
\end{table}
 \\
 The proposed renormalized current is given by
\bea \langle j^{x}\rangle_{ren}&=&q \int_{-\infty}^{\infty} \frac{dk}{2 \pi} \left [\left(k-q A\right) ( |h_k|^2- \frac{1}{\omega^{(0)}_k} ) \right ]\nonumber \\
&= &q \int_{-\infty}^{\infty} \frac{dk}{2 \pi} \left [\left(k-q A\right) |h_k|^2-\frac{\left(k-q A\right)}{\sqrt{(k-qA)^2 +m^2}} \right ]. \eea

Plugging this expression into the semiclassical Maxwell equations \eqref{semimax}, we get the following backreaction equation,
\be -\dot{E}= \ddot{A}=q \int_{-\infty}^{\infty} \frac{dk}{2 \pi}\left [\left(k-q A\right) |h_k|^2-\frac{\left(k-q A\right)}{\sqrt{(k-qA)^2 +m^2}} \right ],\label{bkmax}\ee
together with the equation for the field modes 
\be
\ddot{h}_{k}+\left(m^2+\left(k-qA\right)^2\right)h_{k}=0  \ . \label{equhk2m} \ee

The above semiclassical Maxwell equations are compatible with the conservation of the energy if the renormalized stress-energy tensor is constructed by subtracting until the leading adiabatic order 
\be
\langle T_{00} \rangle_{ren}= \frac{1}{4\pi }\int_{-\infty}^{\infty} dk \left [|\dot{h}_k|^2+ m^2|h_k|^2+(k-q A)^2 |h_k|^2-2\omega^{(0)}_k \right ].\label{scden}
\ee
By direct calculation, one can check that 
\be \partial_\mu \langle T^{\mu\nu} \rangle_{ren} + \partial_\mu T^{\mu\nu}_{elec} =0\label{conservation1}  \ , \ee
where $T_{\mu\nu}^{elec} =   \frac12 E^2 \eta_{\mu\nu}$.

\subsection{Second choice for $\omega^{(0)}_k$}

We can alternatively define 
\be \label{2omega}\omega^{(0)}_k = \sqrt{k^2 + m^2} \equiv \omega \  \ee
and proceed according to the rules of the adiabatic expansion. Here, we assume that $A$ is of adiabatic order $1$. The renormalized current is 
\bea
&&\langle j^{x}\rangle_{ren}=q \int_{-\infty}^{\infty} \frac{dk}{2 \pi } \left [\left(k-q A\right) |h_k|^2-\frac{k}{\omega}+\frac{m^2 q A}{\omega^3} \right ] \label{renjgrav} \ .
\eea
Therefore, the semiclassical Maxwell equations should read as 
\bea
&&-\dot{E}= \ddot{A}=q \int_{-\infty}^{\infty} \frac{dk}{2 \pi }\left [\left(k-q A\right) |h_k|^2-\frac{k}{\omega}+\frac{m^2 q A}{\omega^3} \right ]\label{bkmax}\eea
together with the equation for the field modes, which remain identical to (\ref{equhk2m}). We can also find consistency with the conservation of the energy if it is now defined by subtracting until the second adiabatic order
\be
\langle T_{00} \rangle_{ren}= \frac{1}{4\pi }\int_{-\infty}^{\infty} dk \left [|\dot{h}_k|^2+ m^2|h_k|^2+(k-q A)^2 |h_k|^2 -2\omega + \frac{2 k q A}{\omega} -\frac{m^2 q^2 A^2}{ \omega^3}\right ].\label{scden}
\ee

\subsection{Equivalence between both choices}

In this section, we will study the difference between the two different renormalization schemes that we have on hand.
In order to compare these two prescriptions, we should compare the renormalized scalar currents that we have already obtained, namely 

\be \langle j^{x}\rangle_{ren}^{I}=q \int_{-\infty}^{\infty} \frac{dk}{2 \pi} \left [\left(k-q A\right) |h_k|^2-\frac{\left(k-q A\right)}{\sqrt{(k-qA)^2 +m^2}} \right ], \ee

\bea
\langle j^{x}\rangle_{ren}^{II}=q \int_{-\infty}^{\infty} \frac{dk}{2 \pi } \left [\left(k-q A\right) |h_k|^2-\frac{k}{\omega}+\frac{m^2 q A}{\omega^3} \right ] \label{ren1} \ ,
\eea
where the superindex of the current $\langle j^{x}\rangle_{ren}^{I,II}$ refers to the renormalization recipe that we have used to calculate the electric current ($A$ of adiabatic order $0$ or $1$, respectively). To quantify the difference between the two renormalization notions, it is very convenient to define the following quantity:

\bea
 \bigtriangleup \langle j^{x}\rangle_{ren}=\langle j^{x}\rangle_{ren}^{I}-\langle j^{x}\rangle_{ren}^{II} =q \lim_{\Lambda^{\pm} \to \pm \infty} \int_{\Lambda^-}^{\Lambda^+} \frac{dk}{2 \pi } \left [\frac{k}{\omega}-\frac{m^2 q A}{\omega^3}  - \frac{\left(k-q A\right)}{\sqrt{(k-qA)^2 +m^2}} \right ] \ .
\eea

It is worth it to mention that, even though the subtraction terms are divergent, the difference between them should be finite and the quantity defined above is totally consistent.  It is immediate to see that 
\bea
 \bigtriangleup \langle j^{x}\rangle_{ren}=0 \ .
\eea
That is, in Minkowski space-time the electric current computed by using this two different choices is completely equivalent. This is not an accident of working in two space-time dimensions. We have checked that the same equivalence happens in the four-dimensional theory, although the mathematical expressions are much more involved.  We prefer to maintain the discussion in the simplest two-dimensional case to emphasize the conceptual aspects of the discussion.\\

Note that the most relevant observable in the pair production is the electric current $\langle j^{\mu}\rangle_{ren}$ which serves as a source for an electric field \eqref{semimax}. Because of the equivalence between the two methods, the regularization choice is invisible to the particle production phenomena in Minkowski space-time. We find the same equivalence for the stress-energy tensor $\langle T^{\mu\nu}\rangle_{ren}$.\\


For now, we have obtained two different prescriptions which, at least in absence of gravity, seem to be equally valid. However, this conclusion is no longer valid as soon as gravity is incorporated into the game.

\section{Role of gravity}

As we have already said, although the above two possible choices for $\omega^{(0)}_k$ yield to two equivalent semiclassical forms of the Maxwell equations in Minkowski space, the equivalence does not survive when gravity is present. \\

Let us assume now that our space-time is a two-dimensional expanding universe with metric $ds^2= dt^2 - a^2(t) dx^2$. In this case, the prescription for $a(t)$ is to fix it at order 0. Having this restriction, it appears naturally and dimensionally compatible that the previous adiabatic assignment for $A(t)$ should shift from $0$ to $1$. That is, we should have a hierarchy between the two backgrounds fields.  The leading order corresponds to gravity, and therefore one should replace the definition (\ref{2omega}) by 
\be \label{3omega}\omega^{(0)}_k = \sqrt{k^2/a^2+ m^2}\equiv \omega \ , \ee
instead of the naive generalization 
\be \label{naive}\omega^{(0)}_k = \sqrt{(k-qA)^2/a^2  + m^2} \ . \ee This point has been overlooked in the literature, as recently stressed in Ref. \cite{FN}.
 The gauge field should enter at the next to leading order in the adiabatic expansion: $A(t)$ should be treated as a field of adiabatic order $1$, in the same footing as $\dot a(t)$, as displayed in Table \ref{table3}.  
 
 \begin {table}[h]
 \begin{tabular}{|l|c|}
    \hline
       Field &  \small {Adiabatic order assignment}   \\ \hline
    $ a(t) $ &  0      \\ \hline
    $ \dot a(t)$, $A(t)$ &  1  \\ \hline
   $\ddot a(t)$, $\dot a^2(t)$, $A^2(t)$, $\dot a(t) A(t)$  &2  \\ \hline
    $ \dddot a(t) $, $\ddot a(t) \dot a(t)$, $A^3(t)$, $\dot a(t) \dot A(t)$, ... & 3  \\ \hline
    $ \ddddot a(t) $, ... &  4  \\ \hline    \end{tabular}
\caption{\small{We  summarize the adiabatic order assignment for different numbers of derivatives for the metric and the gauge field. }}
\label{table3} 
\end{table}

\begin {table}[h]

  \begin{tabular}{|c|c|c|c|}
    \hline
  \small {Order of  adiabatic subtractions}      &  $D=4$  & $D=3$ & $D=2$  \\ \hline
    $\langle j^\mu \rangle $ & 3  & 2 & 1 \\ \hline  
    $ \langle T^{\mu\nu} \rangle$ & 4  & 3& 2    \\ \hline
    \end{tabular}
\caption{\small{We  summarize the order of the required adiabatic subtractions for the renormalization of  vacuum expectation values of the most relevant operators, in spacetime dimensions $D=2, 3, 4$. The order corresponds to the scaling dimension of the operators $j^\mu$ and $T^{\mu\nu}$.  }}
\label{table4}
\end{table}

In short, when gravity is present, the rules for the adiabatic subtraction terms are univocally defined, as given in Table III,  and the only consistent adiabatic order assignment for $A(t)$ is $1$.
%
Therefore, the renormalized expression for the electric current should be
\bea
&&\langle j^{x}\rangle_{ren}=q \int_{-\infty}^{\infty} \frac{dk}{2 \pi a^3} \left [\left(k-q A\right) |h_k|^2-\frac{k}{\omega}+\frac{m^2 q A}{\omega^3} \right ] \label{renjgrav} \ .
\eea
Consequently, the semiclassical Maxwell equations in the presence of gravity should read as 

\bea
&&-\dot{E}= \frac{\ddot{A}}{a}-\frac{\dot{A}}{a}\frac{\dot{a}}{a}=q \int_{-\infty}^{\infty} \frac{dk}{2 \pi a^2}\left [\left(k-q A\right) |h_k|^2-\frac{k}{\omega}+\frac{m^2 q A}{\omega^3} \right ]\label{bkmax}\eea
together with the equation for the field modes \eqref{equhk2}.\\

As in Minkowski space-time, and since these equations describe the energy exchange between the classical electric field and the created charged particles, one should expect energy-momentum conservation, namely
\be \nabla_\mu \langle T^{\mu\nu} \rangle_{ren} + \nabla_\mu T^{\mu\nu}_{elec} =0\label{conserv1}  \ , \ee
where $T_{\mu\nu}^{elec} =   \frac12 E^2 g_{\mu\nu}$.
\\ 

After the required adiabatic subtractions, one can obtain the following expression for $\langle T_{00} \rangle_{ren}$:

 \bea
&&\langle T_{00} \rangle_{ren}= \frac{1}{4\pi a}\int_{-\infty}^{\infty} dk \left [|\dot{h}_k-\frac{\dot{a}}{2a}h_k|^2+ m^2|h_k|^2+\frac{1}{a^2}(k-q A)^2 |h_k|^2-2 \omega+\frac{2 k q A}{a^2\omega}-\frac{m^4 \dot a^2}{4 a^2 \omega ^5}-\frac{m^2 q^2 A^2}{a^2 \omega^3} \right].
\label{scden}
\eea
For the $\langle T_{11}\rangle$ component of the stress-energy tensor, we find
\bea
&&\langle T_{11} \rangle_{ren}=\frac{a}{4\pi }\int_{-\infty}^{\infty} dk   \left[ |\dot{h}_k-\frac{\dot{a}}{2a}h_k|^2-  m^2|h_k|^2+\frac{1}{a^2}(k-q A)^2 |h_k|^2 -\langle T^{(0-2)}_{11} \rangle_k\right],\ \label{scT11}\  \eea
where 
\bea
&&\langle T^{(0-2)}_{11} \rangle_k=\frac{2k^2}{a^2\omega}-\frac{2kqA}{a^2\omega} -\frac{2km^2qA}{a^2\omega^3} -\frac{m^4 \ddot a}{2 \omega ^5 a}+\frac{5 m^6 \dot a^2}{4 \omega ^7 a^2}-\frac{3 m^4 \dot a^2}{4 \omega ^5 a^2}+\frac{3 m^4 q^2 A^2}{\omega ^5
   a^2}-\frac{m^2 q^2 A^2}{\omega ^3 a^2}.\nonumber 
 \eea \\

The conservation equation for the zeroth component can be decomposed as
\bea
\nabla_{\mu}\langle T^{\mu0}\rangle_{ren}+ \nabla_\mu T^{\mu0}_{elec}= \partial_0\langle T_{00}\rangle_{ren}+\frac{\dot a}{a}\langle T_{00}\rangle_{ren}+\frac{\dot a}{a^3}\langle T_{11}\rangle_{ren}+ \partial_0 T_{00}^{elec}=0.\label{conservacion2} \
\eea
Plugging  \eqref{scden} and  \eqref{scT11} into \eqref{conservacion2} and using the equation of motion of the scalar field \eqref{equhk2} we get
\bea
\nabla_{\mu}\langle T^{\mu0}\rangle_{ren}+ \nabla_\mu T^{\mu0}_{elec}=\frac{ \dot{A}}{a}\left(\frac{ \ddot{A}}{a}-\frac{ \dot{A}\dot{a}}{a^2}-q \int_{-\infty}^{\infty} \frac{dk}{2 \pi a^2}\left [ \left(k-q A\right)|h_k|^2-\frac{k}{\omega}+\frac{m^2 q A}{\omega^3}\right ]\right)=0 \label {conservation3} \ ,
\eea
where, remarkably, the factor in parentheses is precisely the same appearing in the semiclassical Maxwell equation for the electric field \eqref{bkmax}. A similar result is trivially obtained for 
the remaining component: $\nabla_{\mu}\langle T^{\mu1}\rangle_{ren}+ \nabla_\mu T^{\mu1}_{elec}=0$.
 Therefore, the semiclassical Maxwell equations must be satisfied to ensure energy-momentum conservation. \\

If one assumes  that $A(t)$ is of adiabatic order $0$, the energy-momentum conservation does not hold anymore. With this alternative adiabatic assignment, and renormalizing $\langle T^{\mu\nu}\rangle$ up to and including the second adiabatic order, one obtains 
\bea
\nabla_{\mu}\langle T^{\mu0}\rangle_{ren}+ \nabla_\mu T^{\mu0}_{elec}\neq 0 \ . \eea
In this case, the left-hand side in the above equation is proportional to  $E \langle j^x\rangle^{I,(2)}$, where $\langle j^x\rangle^{I,(2)}$ is the second adiabatic order of the electric current, which cannot be properly absorbed into the renormalization substractions of the electric current (see Table \ref{table4} ). We note that another inconsistency is the disagreement with the trace anomaly \cite{FN} in the massless limit.

\section{Conclusions}\label{conclusions}

The main lesson of this paper is clear. Gravity plays a fundamental role in fixing unambiguously the adiabatic rules in the renormalization of the source operators and hence in the correct expression for the backreaction equations. The basic consistency check is the exact compatibility of the backreaction equations with the covariant conservation of the total stress-energy tensor. The result (\ref{conservation3}) can be regarded as a theorem, and the crucial hypotheses are the adiabatic order assignments proposed in Table \ref{table3} and the renormalization prescriptions given in Table \ref{table4}. The adiabatic order assignment $0$ for the gauge field $A(t)$, with the proposal (\ref{naive}), is in sharp tension with  energy-momentum conservation. However,  
in Minkowski space, the two distinct notions for constructing  the adiabatic subtraction terms turn out to be equivalent, as shown explicitly in Sec. 3. This may explain why the inconsistency of (\ref{naive}) has been largely unnoticed.\\

\section*{Acknowledgments}

 This work has been supported by the Spanish MINECO research Grants No. FIS2014-57387-C3-1-P and No.   FIS2017-84440-C2-1-P; Project No. SEJI/2017/042 (Generalitat Valenciana); the COST action CA15117 (CANTATA), supported by COST (European Cooperation in Science and Technology); and the Severo Ochoa Program, Grant No.  SEV-2014-0398.    A. F. is supported by the Severo Ochoa Ph.D. fellowship, Grant No.  SEV-2014-0398-16-1 and the European Social Fund. S. Pla is supported by the Formaci\'on del Personal Universitario Ph.D. fellowship, Grant No. FPU16/05287. 

\end{document}